\documentclass[twocolumn]{IEEEtran}
\usepackage{amsmath}
\usepackage{amssymb,amsthm}
\usepackage{graphicx}
\usepackage{psfrag}
\usepackage{epsfig}
\usepackage{subfigure}
\usepackage{color}
\usepackage{latexsym, amssymb, array, multirow}

\usepackage{bbm}
\input{mathlig}

\newcommand{\muspace}{\mspace{1mu}}

\DeclareRobustCommand{\scond}{\mathchoice{\muspace\vert\muspace}{\vert}{\vert}{\vert}}
\mathlig{|}{\scond}

\DeclareRobustCommand{\discint}{\mathchoice{\mspace{-1.5mu}:\mspace{-1.5mu}}{\mspace{-1.5mu}:\mspace{-1.5mu}}{:}{:}}
\mathlig{::}{\discint}

\def\tr{\mathop{\rm tr}\nolimits}%
\def\Cov{\mathop{\rm Cov}\nolimits}%
\newcommand{\Xv}{{\bf X}}
\newcommand{\Yv}{{\bf Y}}
\newcommand{\Zv}{{\bf Z}}
\newcommand{\Uv}{{\bf U}}

\newcommand{\xv}{{\bf x}}
\newcommand{\yv}{{\bf y}}

\newcommand{\uv}{{\bf u}}

\newcommand{\mh}{{\hat{m}}}

\DeclareMathOperator\E{\textsf{E}}

\def\textiid{i.i.d.\@\xspace}
\newcommand\iid{\ifmmode\text{ i.i.d. } \else \textiid \fi}

\newcommand{\sfrac}[2]{\mbox{\small$\displaystyle\frac{#1}{#2}$}}
\newcommand{\half}{\sfrac{1}{2}}

\def\mathllap{\mathpalette\mathllapinternal}
\def\mathllapinternal#1#2{%
  \llap{$\mathsurround=0pt#1{#2}$}}

\def\clap#1{\hbox to 0pt{\hss#1\hss}}
\def\mathclap{\mathpalette\mathclapinternal}
\def\mathclapinternal#1#2{%
  \clap{$\mathsurround=0pt#1{#2}$}}

\let\oldstackrel\stackrel
\renewcommand{\stackrel}[2]{\oldstackrel{\mathclap{#1}}{#2}}

\renewcommand{\hbar}{h\mathllap{\overline{\vphantom{h}\hphantom{\rule{4.6pt}{0pt}}}\mspace{0.77mu}}}

\catcode`~=11 
\newcommand{\urltilde}{\kern -.06em\lower -.06em\hbox{~}\kern .02em}
\catcode`~=13 

\newtheorem{lemma}{Lemma}
\newtheorem{theorem}{Theorem}
\newtheorem{proposition}{Proposition}
\theoremstyle{definition}

\theoremstyle{remark}
\newtheorem{remark}{Remark}

\newcommand{\Rcs}{R_\mathrm{CS}}
\newcommand{\Rpdf}{R_\mathrm{PDF}}
\newcommand{\Rnpdf}{R_\mathrm{NPDF}}
\newcommand{\Rcf}{R_\mathrm{CF}}
\newcommand{\Rdf}{R_\mathrm{DF}}
\newcommand{\Rdt}{R_\mathrm{DT}}
\newcommand{\Dpdf}{\Delta_\mathrm{PDF}}
\newcommand{\Dnpdf}{\Delta_\mathrm{NPDF}}
\newcommand{\Dcf}{\Delta_\mathrm{CF}}
\newcommand{\hermitian}{\scalebox{0.9}{\scriptsize${}\mathsf{H}$}}
\newcommand{\mystar}{*}

\begin{document}
\title{The Approximate Capacity\\
of the MIMO Relay Channel}

\author{Xianglan Jin and Young-Han Kim%
\thanks{This work was presented in part at the 2014 IEEE International Symposium on 
Information Theory, Honolulu, Hawaii.}%
\thanks{X.~Jin is with the Department of Electrical and Computer Engineering, Pusan National University, Busan 609-735, Korea (e-mail: jinxl77@pusan.ac.kr) and is currently visiting the Department of Electrical and Computer Engineering,
University of California, San Diego, La Jolla, CA 92093-0407, USA.}
\thanks{Y.-H.~Kim is with the Department of Electrical and Computer Engineering,
University of California, San Diego,
La Jolla, CA 92093-0407, USA (e-mail:yhk@ucsd.edu).}
}

\maketitle

\begin{abstract}
Capacity bounds are studied for the multiple-antenna complex Gaussian relay channel
with $t_1$ transmitting antennas at the sender, $r_2$ receiving
and $t_2$ transmitting antennas at the relay, and $r_3$ receiving antennas at the receiver.
It is shown that the partial decode--forward coding scheme achieves within
$\min(t_1,r_2)$ bits from the cutset bound and at least one half of the cutset bound,
establishing a good approximate expression of the capacity.
A similar additive gap of 
$\min(t_1  + t_2, r_3) + r_2$ bits is shown to be
achieved by the compress--forward coding scheme.
\end{abstract}

\IEEEpeerreviewmaketitle

\section{Introduction}

The relay channel, whereby point-to-point communication between a sender
and a receiver is aided by a relay,
is a canonical building block for cooperative wireless communication. Introduced by van der Meulen~\cite{van-der-Meulen1971b}, this channel model has been studied extensively
in the literature, including the now classical paper by Cover and El Gamal~\cite{Cover--El-Gamal1979}. 
The problem of characterizing the capacity in a computable expression, however,
remains open even for simple channel models, and consequently a large body of the literature has been devoted to the study of upper and lower bounds on the capacity. Reminiscent of the max-flow min-cut
theorem~\cite{Ford--Fulkerson1956}, the cutset bound was established by Cover and El Gamal~\cite{Cover--El-Gamal1979}, which sets an intuitive upper limit on the capacity.
On the other direction, there are a myriad of coding
schemes, typically referred to as ``*--forward''~\cite{Kim2014}, 
each establishing a lower bound on the capacity.
Among these, the two most versatile coding schemes are
partial decode--forward~\cite[Th.~7]{Cover--El-Gamal1979} and compress--forward~\cite[Th.~6]{Cover--El-Gamal1979},
which are complementary to each other (providing digital-to-digital and analog-to-digital relays, respectively)
and have been successfully extended to
general relay networks for unicast, multicast, broadcast, and multiple access 
\cite{Kramer--Gastpar--Gupta2005, Yassaee--Aref2011, Lim--Kim--El-Gamal--Chung2011, Lim--Kim--Kim2015}.

The Gaussian relay channel, whereby the signals from the sender and the relay are corrupted by additive white Gaussian noise, is one of the most basic channel models studied in the literature.
The capacity of the Gaussian relay channel, however, is again unknown for any nondegenerate channel parameter.
Instead, the following results have been established for single-antenna Gaussian relay channels.
\begin{itemize}
\item Partial decode--forward and compress--forward,
respectively, achieve within one bit from the cutset bound~\cite{Chang--Chung--Lee2010, Avestimehr--Diggavi--Tse2011}.

\item Partial decode--forward, which is a superposition of decode--forward and direct transmission, reduces to the better of the two~\cite{El-Gamal--Mohseni--Zahedi2006}. 
\end{itemize}
These results establish simple approximate expressions of the capacity,
which are particularly useful in high signal-to-noise ratio (SNR).
A natural question arises on how these results can be extended to multiple-antenna (also known as multiple-input 
multiple-output or MIMO) Gaussian relay channels. 

Capacity bounds for MIMO relay channels have been studied in several papers.
By convex optimization techniques \cite{Boyd--Vandenberghe2004}, Wang, Zhang, and
H\o{}st-Madsen~\cite{Wang--Zhang--Host-Madsen2005} derived upper and lower bounds based
on looser versions of the cutset bound and the decode--forward bound. These results have been
improved by more advanced coding schemes (partial decode--forward and compress--forward) 
with suboptimal decoding rules by Simoens, Munoz-Medina, Vidal, and 
del Coso~\cite{Simoens--Munoz-Medina--Vidal--del-Coso2009}
and Ng and Foschini~\cite{Ng--Foschini2011}.
The usual focus of this line of work, however, has been on the optimization of resources (power and
bandwidth) for practical implementations and on numerical computation of resulting capacity bounds 
(see also \cite{Gerdes--Weiland--Utschick2013}).
The most relevant to our main question is a recent result by Kolte, \"Ozg\"ur, and El~Gamal \cite{Kolte--Ozgur--El-Gamal2015} on a general MIMO relay network, which carefully compares the noisy network coding lower bound for the general unicast relay network \cite{Lim--Kim--El-Gamal--Chung2011} with the cutset bound, which can be readily specialized to the 3-node relay channel. In the same vein, another recent study
by Gerdes, Hellings, Weiland, and Utschick \cite{Gerdes--Hellings--Weiland--Utschick2014} establishes
the optimal input distribution of the partial decode--forward lower bound for the MIMO relay channel, the corresponding result
of which for the single-antenna case is immediate since partial decode--forward is the better of decode--forward and direct transmission.

This paper provides more direct and comprehensive answers to our main question through an elementary yet careful analysis
of the partial decode--forward and compress--forward lower bounds for the MIMO relay channel.
The main contributions are summarized as follows.
 
\begin{itemize}
\item For the complex Gaussian relay channel with $t_1$ transmitting antennas at the sender, $r_2$ receiving and $t_2$ transmitting antennas at the relay, and $r_3$ receiving antennas at the receiver, we show that the partial decode--forward achieves within $\min (t_1,r_2)$ bits of the cutset bound (Theorem~\ref{thm:gap-pdf}). 

\item This gap is somewhat relaxed when noncoherent transmission is employed (Proposition~\ref{prop:gap-npdf}).

\item Unlike the single-antenna counterpart, partial decode--forward can achieve rates arbitrarily higher than the better of decode--forward and direct transmission in MIMO relay channels (Proposition~\ref{prop:gap-df}).

\item To complement the additive gap result, we show that both coherent and noncoherent partial decode--forward coding schemes achieve at least half the cutset bound (Theorem~\ref{thm:multiplicative-gap-pdf}).

\item We show that compress--forward achieves $\min (t_1  + t_2,r_3) + r_2$ bits within the cutset bound (Theorem~\ref{thm:gap-cf}).

\item We establish similar results for half-duplex relay channel models \cite{El-Gamal--Mohseni--Zahedi2006}, \cite{Host-Madsen--Zhang2005}, \cite{Liang--Veeravalli2005} (Section~\ref{sec:half-duplex}).
\end{itemize}
In conclusion, the paper establishes simple approximate expressions of the capacity,
which are particularly useful in high and low SNR. Beyond these analytical results,
we also discuss how these expressions
can be computed efficiently.

The rest of the paper is organized as follows. 
In the next section, we formally define the channel model and
review the cutset upper bound, the
partial decode--forward lower bound, and the compress--forward lower bound on the capacity. The main results on
additive and multiplicative gaps for partial decode--forward and compress--forward are also stated therein. 
The proofs of these results are given in Sections~\ref{sec:pdf} and~\ref{sec:cf}.
Section~\ref{sec:computation-capacity} is devoted to the computational aspects of our results, namely, how the capacity
bounds can be computed efficiently via appropriate convex optimization formulations. Using these computational tools, the main results are verified by numerical simulations.
In Section \ref{sec:half-duplex}, half-duplex MIMO relay channels are discussed.

Throughout the paper, we use the following notation.
The superscript $(\cdot)^{\hermitian}$ denotes the complex conjugate transpose of a (complex) matrix;
$\tr(\cdot)$ denotes the trace of a matrix; 
$I_n$ denotes the $n \times n$ identity matrix;
$\mathbb{C}^{n\times m}$ denotes a set of $n\times m$ complex
matrices; $A\succeq B$ denotes that $A-B$ is hermitian and positive semidefinite;
$\E(\cdot)$ denotes the expectation with respect to the random variables in the argument.

\section{Problem Setup and Main Results}\label{sec:problem-setup-main-results}

\begin{figure}[h]
\centering
\small
\bigskip
\psfrag{x}[r]{$\Xv_1$}
\psfrag{a}[b]{$G_{21}$}
\psfrag{z1}[b]{$\Zv_2$}
\psfrag{y1}[l]{$\Yv_{\!2}$}
\psfrag{x1}[r]{$:\hspace{-2pt}\Xv_2$\!}
\psfrag{b}[b]{$G_{32}$}
\psfrag{z}[b]{$\Zv_3$}
\psfrag{y}[l]{$\Yv_3$}
\psfrag{1}[t]{$G_{31}$}
\includegraphics[scale=0.42]{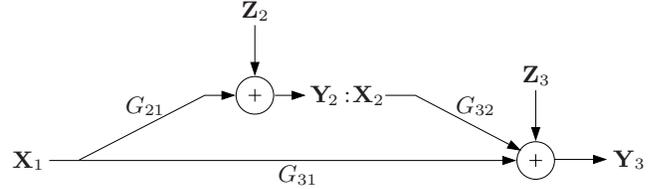}
\caption{The MIMO relay channel.}
\label{fig:relay}
\end{figure}

We model the point-to-point communication system with a relay
as a MIMO relay channel 
with sender node~1, relay node~2, and receiver node~3; see Fig.~\ref{fig:relay}.
Throughout the paper, we assume the complex signal model, but corresponding results
for the real case can be easily obtained; see the conference version \cite{Jin--Kim2014} of the current paper
for some results on the real model.
The relay and the receiver have $r_2$ and $r_3$ receiving antennas with respective channel outputs
\begin{equation} \label{eq:channel-model}
\begin{split}
	\Yv_2 &= G_{21} \Xv_1 + \Zv_2,\\
	\Yv_3 &= G_{31} \Xv_1 + G_{32} \Xv_2 + \Zv_3,
\end{split}
\end{equation}
where $G_{21}\in {\mathbb{C}^{r_2\times t_1}}$, $G_{31}\in {\mathbb{C}^{r_3\times t_1}}$, and $G_{32}\in {\mathbb{C}^{r_3\times t_2}}$ are complex channel gain matrices, and $\Zv_2 \sim \mathrm{CN}(0, I_{r_2})$ and $\Zv_3 \sim {\mathrm{CN}}(0, I_{r_3})$ are independent complex Gaussian noise components.
For simplicity, we will often use the shorthand notation
\begin{align*}
G_{3\mystar} = \begin{bmatrix} G_{31} &G_{32} \end{bmatrix} \quad\text{and}\quad
G_{\mystar 1} = \begin{bmatrix}
G_{21} \\
G_{31}
\end{bmatrix}.
\end{align*}
We assume that the sender and the relay have $t_1$ and $t_2$ transmitting antennas, respectively, with average power constraint $P$.
As in the standard relay channel model~\cite{Cover--El-Gamal1979}, the encoder is defined by $\xv_1^n(m)$,
the relay encoder is defined by $\xv_{2i}(\yv_2^{i-1})$, $i = 1,\ldots,n$, and the decoder is defined by $\mh(\yv_3^n)$. 
We assume that the message $M$ is uniformly distributed over the message set. The
average probability of error is defined as $P_e^{(n)}= P\{ ̂\hat{M}\ne M\}$. A rate $R$ is said to be achievable
for the relay channel if there exists a sequence of $(2^{nR}, n)$ codes such that $\lim_{n\to\infty} P_e^{(n)}=0$. The
capacity $C$ of the relay channel is the supremum of all achievable rates. 

The following upper bound on the capacity is well known.
\begin{proposition}[{Cutset bound~\cite[Th.~4]{Cover--El-Gamal1979}}]
The capacity $C$ of the MIMO relay channel is upper bounded by
\begin{align}
\Rcs 
&= \sup_{F(\xv_1,\xv_2)} \min \bigl\{ 
  I(\Xv_1, \Xv_2; \Yv_3), \notag\\[-0.75em]
&\hphantom{\;\mathrel{=} \sup_{F(\xv_1,\xv_2)} \min \bigl\{ } 
  I(\Xv_1;\Yv_2,\Yv_3|\Xv_2)\bigr\} \label{eq:cs-1} \\
&= \max_{K} \min  \bigl\{ {}
  \log |I_{r_3}  +  G_{3\mystar} K G_{3\mystar}^{\hermitian} |, \notag\\[-0.5em]
&\hphantom{\;\mathrel{=} \max_{K} \min \bigl\{ }
  \log | I_{r_2+r_3}  +  G_{\mystar 1} K_{1|2} G_{\mystar 1}^{\hermitian} | \bigr\} \label{eq:cs-2} \\
&= \max_{K} \min  \bigl\{ {}
  \log |I_{r_2}  +  [G_{31}~G_{32}] K [G_{31}~G_{32}]^{\hermitian} |, \notag\\[-0.5em]
&\hphantom{\;\mathrel{=} \max_{K} \min \bigl\{}
  \log | I_{t_1}  +  (G_{ 21}^{\hermitian} G_{ 21} +  G_{ 31}^{\hermitian} G_{ 31})  K_{1|2} | \bigr\} \label{eq:cs-3}
\end{align}
where the supremum in~\eqref{eq:cs-1} is over all joint distributions $F(\xv_1,\xv_2)$ 
such that $\E(\Xv_j^{\hermitian}\Xv_j) \le P$, $j = 1,2$,
the maxima in~\eqref{eq:cs-2} and~\eqref{eq:cs-3} are over all $(t_1+t_2) \times (t_1+t_2)$ matrices
\begin{equation} \label{eq:covariance}
K = 
\begin{bmatrix}
K_1 & K_{12} \\
K_{12}^{\hermitian} & K_2
\end{bmatrix}
\succeq 0
\end{equation}
such that $\tr(K_j) \le P$, $j = 1,2$, 
and 
\[
K_{1|2} = K_1 - K_{12} K_2^{-1} K_{12}^{\hermitian}.
\] 
\end{proposition}

The equality in~\eqref{eq:cs-3} is justified by the following fact that will be used repeatedly throughout the paper.

\begin{lemma}\label{lemm:det-inequality}
For $\gamma \in [0,1]$, $r \times t$ matrix $G$, and $t \times t$ matrix $K \succeq 0$,
\begin{align} 
| I_r + \gamma G K G^{\hermitian} | 
&= |I_t + \gamma G^{\hermitian} G K | \notag\\
&\ge \gamma^{\min(t,r)} |I_r + G K G^{\hermitian}|.
\label{eq:matrix}
\end{align}
\end{lemma}

We compare the cutset bound with two lower bounds on the capacity.
The first lower bound is based on the partial decode--forward coding scheme, in which the relay recovers part of the message
and forwards it.

\begin{proposition}[Partial decode--forward bound
{\cite[Th.~7]{Cover--El-Gamal1979}}]

\label{prop:pdf}

The capacity $C$ of the MIMO relay channel is lower bounded by
\begin{align}
\Rpdf
&= \sup\min\bigl\{ {}
    I(\Xv_1, \Xv_2; \Yv_3), \notag\\
&\hphantom{\;\mathrel{=} \sup\min\bigl\{ {} } 
	I(\Uv; \Yv_2 | \Xv_2) + I(\Xv_1;\Yv_3|\Xv_2, \Uv)\big\} \label{eq:pdf-1}\\
&=\sup\min\bigl\{ {}
    I(\Xv_1, \Xv_2; \Yv_3),  \notag\\
&\hphantom{\;\mathrel{=} \sup\min\bigl\{ {} } 
	I(\Xv_1;\Uv, \Yv_3|\Xv_2)-I( \Xv_1;\Uv | \Xv_2,\Yv_2) \big\} \label{eq:pdf-2}
\end{align}
where the suprema are over all joint distributions $F(\uv,\xv_1,\xv_2)$ such that
$\E(\Xv_j^{\hermitian}\Xv_j) \le P$, $j = 1,2$.
\end{proposition}

\begin{remark}
The partial decode--forward lower bound does not increase
by (coded) time sharing.
\end{remark}

The partial decode--forward lower bound can be relaxed in several directions. First, by 
limiting the input distribution to a more practical product form, we
obtain the \emph{noncoherent} partial decode--forward lower bound:
\begin{align}
\Rnpdf
&= \sup\min\bigl\{ {}
    I(\Xv_1, \Xv_2; \Yv_3), \notag\\
&\hphantom{\;\mathrel{=} \sup\min\bigl\{ {} } 
	I(\Uv; \Yv_2 | \Xv_2) + I(\Xv_1;\Yv_3|\Xv_2, \Uv)\big\} \label{eq:npdf-1}\\
&=\sup\min\bigl\{ {}
    I(\Xv_1, \Xv_2; \Yv_3),  \notag\\
&\hphantom{\;\mathrel{=} \sup\min\bigl\{ {} } 
	I(\Xv_1;\Uv, \Yv_3|\Xv_2)-I( \Xv_1;\Uv | \Xv_2,\Yv_2) \big\} \label{eq:npdf-2}
\end{align}
where the suprema are over all \emph{product} distributions $F(\uv,\xv_1)F(\xv_2)$ such that
$\E(\Xv_j^{\hermitian}\Xv_j) \le P$, $j = 1,2$.
Second, by setting $\Uv = \Xv_1$, which is equivalent to having the relay recover the entire message,
we obtain the decode--forward lower bound:
\begin{align}
\Rdf
&= \sup \min\bigl\{ I(\Xv_1, \Xv_2; \Yv_3),\, I(\Xv_1; \Yv_2 | \Xv_2) \big\} \label{eq:df-1} \\
&= \max_{K}\min  \bigl\{ {}
	\log |I_{r_3}  +  G_{3\mystar} K G_{3\mystar}^{\hermitian}  |, \notag\\[-0.5em]
&\hphantom{\;\mathrel{=} \max_{K}\min  \bigl\{ {}}
	\log  | I_{r_2}  +  G_{21} K_{1|2} G_{21}^{\hermitian} |\bigr\} \label{eq:df-2}
\end{align}
where the supremum in~\eqref{eq:df-1} is over all distributions $F(\xv_1, \xv_2)$ such that
$\E(\Xv_j^{\hermitian}\Xv_j) \le P$, $j = 1,2$, and the maximum in~\eqref{eq:df-2} is over 
all $(t_1+t_2) \times (t_1+t_2)$ matrices
$K\succeq 0$ of the form~\eqref{eq:covariance} such that $\tr(K_j) \le  P$, $j = 1,2$.
Third, by setting $\Uv = \emptyset$ and $\Xv_2 = 0$, we obtain the direct-transmission lower bound:
\begin{align}
\Rdt
&= \sup I(\Xv_1; \Yv_3) \notag \\
&= \max_{K_1} 
	 \log |I_{r_3}  +  G_{31} K_1 G_{31}^{\hermitian} |
 \label{eq:dt}
\end{align}
where the supremum is over all distributions $F(\xv_1)$ such that
$\E(\Xv_1^{\hermitian}\Xv_1) \le P$ and the maximum is over all $t_1\times t_1$ matrices $K_1 \succeq 0$.

\begin{remark}
Since decode--forward and direct transmission schemes are two special cases of partial decode--forward, we have in general
\begin{equation} \label{eq:pdf-df-dt}
\Rpdf\ge \max (\Rdf,\Rdt).
\end{equation}
\end{remark}

Next, we present another important lower bound, in which the relay compresses its noisy observation
instead of recovering the message.

\begin{proposition}[Compress--forward bound {\cite[Th.~6]{Cover--El-Gamal1979}, \cite{El-Gamal--Mohseni--Zahedi2006}}]\label{Prop:CF-LB}
The capacity $C$ of the MIMO relay channel is lower bounded by
\begin{align}
\Rcf
&= \sup
I(\Xv_1;\hat{\Yv}_2,\Yv_3|\Xv_2) \label{eq:cf-1}
\end{align}
where the supremum is over all conditional distributions $F(\xv_1)F(\xv_2)  F(\hat{\yv}_2 |\yv_2 ,\xv_2)$ such that 
$\E(\Xv_j^{\hermitian}\Xv_j) \le P$, $j = 1,2$ and 
\[
I(\Xv_2;\Yv_3)\ge I(\Yv_2;\hat{\Yv}_2|\Xv_2,\Yv_3).
\]
This lower bound can be expressed equivalently as
\begin{align}
\Rcf
&=\sup\min \bigl\{ {}
	I(\Xv_1 , \Xv_2; \Yv_3) - I(\Yv_2;\hat\Yv_2|\Xv_1 ,\Xv_2,\Yv_3),\notag\\
&\hphantom{\;\mathrel{=} \sup\min \bigl\{ {} }
	I(\Xv_1;\hat{\Yv}_2 ,\Yv_3|\Xv_2)\bigr \} \label{eq:cf-2}
\end{align}
where the supremum is over all conditional distributions $F(\xv_1)F(\xv_2)  F(\hat{\yv}_2 |\yv_2 ,\xv_2)$
such that $\E(\Xv_j^{\hermitian}\Xv_j) \le P$, $j = 1,2$.
\end{proposition}

\begin{remark}
The compress--forward lower bound
before taking the supremum in~\eqref{eq:cf-1} or~\eqref{eq:cf-2} is not a convex function of the conditional distribution
$F(\xv_1)F(\xv_2)  F(\hat{\yv}_2 |\yv_2 ,\xv_2)$
in general
and can be potentially improved by (coded) time sharing~\cite[Remark~16.4]{El-Gamal--Kim2011}.
\end{remark}

\begin{remark}
\label{rem:cf-dt}
By setting $\hat\Yv_2 = \emptyset$, compress--forward reduces to direct transmission
and thus $\Rcf \ge \Rdt$.
\end{remark}


We are now ready to state the main results of the paper.

\begin{theorem}\label{thm:gap-pdf}
For every $G_{21}, G_{31}, G_{32}$, and $P$,
\begin{align}
\Dpdf &:= \Rcs - \Rpdf 
\le  \min( t_1,  r_2). \label{eq:gap-pdf}
\end{align} 
\end{theorem}

As a supplement to the additive gap result in Theorem~\ref{thm:gap-pdf}, which is useful in approximating the capacity in high SNR, we establish the following multiplicative gap to provide a tighter approximation in low SNR.

\begin{theorem} \label{thm:multiplicative-gap-pdf}
For every $G_{21}, G_{31}, G_{32}$, and $P$,
\begin{align}
\frac{\Rcs}{\Rpdf} \le 2. \label{eq:multiplicative-gap-pdf}
\end{align}
In other words, partial decode--forward always achieves at least half the capacity.
\end{theorem}

The above results can be relaxed by using the noncoherent partial decode--forward.
\begin{proposition} \label{prop:gap-npdf}
For every $G_{21}, G_{31}, G_{32}$, and $P$,
\begin{align}
\Dnpdf 
&:= \Rcs - \Rnpdf \notag\\
&\le  \max \bigl[\min(t_1,r_2),\, \min(t_1  +  t_2,\, r_3) \bigr] \label{eq:gap-npdf}
\end{align}
and
\begin{equation} \label{eq:multiplicative-gap-npdf}
\frac{\Rcs}{\Rnpdf} \le 2.
\end{equation}
\end{proposition}

For the single-antenna case, 
the partial decode--forward lower bound
can be shown \cite[Sec.~II]{El-Gamal--Mohseni--Zahedi2006} to be \emph{equal} to the maximum of the decode--forward and direct--transmission lower bounds; cf.~\eqref{eq:pdf-df-dt}.
For multiple antennas, however, partial decode--forward is in general much richer than decode--forward and direct transmission.
\begin{proposition} \label{prop:gap-df}
If $t_1,t_2,r_2,r_3 \ge 2$,
\[
\sup_{G_{21}, G_{31}, G_{32}, P} \bigl[ \Rpdf - \max ( \Rdf, \Rdt ) \bigr] = \infty.
\]
\end{proposition}

In~\cite[Th.~1]{Kolte--Ozgur--El-Gamal2015}, Kolte,  \"Ozg\"ur, and El~Gamal derived a capacity lower bound for
general MIMO relay networks. When specialized to the three-node relay network and further tightened,
this result yields the following channel-independent capacity approximation.
 
\begin{proposition}
\label{Prop:cf-gap-Kolte--Ozgur--El-Gamal2014}

For every $G_{21}, G_{31}, G_{32}$, and $P$,
\begin{align}
&\Dcf \notag \\
&:= \Rcs - \Rcf \notag \\
&\le \min_{\sigma^2}  \max \biggl[ 
	\min(t_1  + t_2,\, r_3) \log\bigg(1+\frac{t_1+t_2}{\min(t_1+t_2,r_3)}\bigg)\notag \\
&\hphantom{\,\le \min_{\sigma^2}  \max \biggl[}
	\quad+ r_2 \log (1 + 1/\sigma^2),\notag\\
&\hphantom{\,\le \min_{\sigma^2}  \max \biggl[}
	\min (t_1,\, r_2 + r_3) \log(1 + \sigma^2) \biggr]. \label{eq:cf-gap1-Kolte}
\end{align}
\end{proposition}

We tighten this result further as follows.

\begin{theorem}\label{thm:gap-cf}
For every $G_{21}, G_{31}, G_{32}$, and $P$,
\begin{align}
\Dcf 
&\le \min_{\sigma^2}  \max \bigl[ 
	\min(t_1  + t_2,\, r_3) + r_2 \log (1 + 1/\sigma^2),\notag\\[-0.5em]
&\hphantom{\,\le \min_{\sigma^2}  \max \bigl[}
	\min (t_1,\, r_2 + r_3) \log(1 + \sigma^2) \bigr] \label{eq:cf-gap1}\\
&\le    \min (t_1  + t_2,\, r_3) + r_2.  \label{eq:cf-gap2}
\end{align}
\end{theorem}

No multiplicative gap is known between the compress--forward lower bound and the cutset bound.
This follows partly from the fact that the distribution that attains the suprema in \eqref{eq:cf-1} and \eqref{eq:cf-2}
is rather difficult to characterize. It can be shown, however, that when restricted to Gaussian distributions,
the compress--forward lower bound (even with time sharing) may have an unbounded multiplicative gap from the cutset bound.
As a compromise, we state the following simple consequence of Remark~\ref{rem:cf-dt} and the proof of 
Theorem~\ref{thm:multiplicative-gap-pdf}.

\begin{proposition}
For every $G_{21}, G_{31}, G_{32}$, and $P$,
\begin{align}
\frac{\Rcs}{\max(\Rdf,\Rcf)}\le 2.
\end{align}
\end{proposition}

\section{Partial Decode--Forward}\label{sec:pdf}

In this section, we establish the results on partial decode--forward
stated in the previous section (Theorems~\ref{thm:gap-pdf} and~\ref{thm:multiplicative-gap-pdf},
and Propositions~\ref{prop:gap-npdf} and~\ref{prop:gap-df}).

\subsection{Partial Decode--Forward (Proof of Theorem \ref{thm:gap-pdf})} \label{sec:co-pdf}
We evaluate the partial decode--forward lower bound in~\eqref{eq:pdf-1}
with $(\Xv_1,\Xv_2) \sim \mathrm{CN}(0, K)$, where $K \succeq 0$ is of the form
in~\eqref{eq:covariance}, and
\begin{equation} \label{eq:uv-choice}
\Uv = G_{21}\Xv_1+\Zv'_2,
\end{equation}
where $\Zv'_2\sim\mathrm{CN}(0,I_{r_2})$ is independent of $(\Xv_1, \Xv_2, \Zv_2, \Zv_3)$.
Note that $(\Xv_1, \Xv_2, \Uv, \Yv_3)$ has the same distribution as $(\Xv_1, \Xv_2, \Yv_2, \Yv_3)$.
The first term of the minimum in~\eqref{eq:pdf-1} is
 \begin{equation} \label{eq:pdf-bd-1}
	I(\Xv_1, \Xv_2; \Yv_3) = 
	\log\bigr |I_{r_3}  +  G_{3\mystar} K G_{3\mystar}^{\hermitian} \bigr |.
\end{equation}
For the second term, since 
\begin{align*}
	&\Cov(\Xv_1 | \Uv, \Xv_2) \notag\\
	&=\Cov(\Xv_1 | \Yv_2, \Xv_2)\notag\\
	&= K_{1|2}-K_{1|2}G_{21}^{\hermitian}( I_{r_2}+G_{21}K_{1|2}G_{21}^{\hermitian})^{-1}G_{21}K_{1|2} \notag\\
	&= K_{1|2}\Bigl(I_{t_1}+G_{21}^{\hermitian}G_{21}K_{1|2}\Bigr)^{-1},
\end{align*}
we have 
\begin{align}
&I(\Uv;\Yv_2|\Xv_2)+I(\Xv_1;\Yv_3|\Xv_2,\Uv) \notag \\
&=\log\frac{|I_{r_3}+G_{31}\Cov(\Xv_1|\Uv, \Xv_2) G_{ 31}^{\hermitian}|}
	{|I_{r_2}+G_{21}\Cov(\Xv_1|\Uv,\Xv_2) G_{21}^{\hermitian}|} 	  \notag \\
&\qquad+\log{ |I_{r_2}+G_{21}K_{1|2}G_{21}^{\hermitian}|}\notag\\
&=\log|I_{t_1}+(G_{21}^{\hermitian}G_{21}+G_{31}^{\hermitian}G_{31})K_{1|2}| \notag \\
&\qquad+\log  \frac{ |I_{t_1}+ G_{21}^{\hermitian}G_{ 21}K_{1|2}|}
	         {|I_{t_1} + 2G_ { 21}^{\hermitian}G_{21}K_{1|2}|}
			 \label{eq:pdf-bd-2} \\
&\ge \log|I_{t_1}+(G_{21}^{\hermitian}G_{21}+G_{31}^{\hermitian}G_{31})K_{1|2}| 
      -\min(t_1,r_2)   \label{eq:pdf-bd-3}
\end{align}
where the last inequality follows by Lemma~\ref{lemm:det-inequality}.
Comparing~\eqref{eq:pdf-bd-1} and~\eqref{eq:pdf-bd-3} with the cutset bound in~\eqref{eq:cs-3}
completes the proof of Theorem~\ref{thm:gap-pdf}.

We can prove Theorem~\ref{thm:gap-pdf} alternatively using the following result that is applicable 
to a more general class of relay channels and follows by 
setting $p(u|x_1,x_2) = p_{Y_2|X_1,X_2}(u|x_1,x_2)$
in the second form of the partial decode--forward lower bound in~\eqref{eq:pdf-2}.

\begin{proposition}\label{prop:gap-pdf-general}
For a discrete memoryless relay channel $p(y_2,y_3|x_1,x_2) = p(y_2|x_1,x_2)p(y_3|x_1,x_2)$,
\begin{align*}
\Dpdf &:=\Rcs-\Rpdf\\
&\le \max_{p(x_1,x_2)} I(X_1;U| X_2,Y_2)
\end{align*}
where $p(u|x_1,x_2) = p_{Y_2|X_1,X_2}(u|x_1,x_2)$.
\end{proposition}

Now, applying Proposition~\ref{prop:gap-pdf-general} to the MIMO relay channel,
we have 
\begin{align*}
	&I( \Xv_1;\Uv| \Xv_2,\Yv_2)\notag\\
	&=\log \bigl|I_{r_3}+ G_{21}\Cov(\Xv_1|\Yv_2,\Xv_2) G_{21}^{\hermitian}\bigr|\notag\\
	&=\log\bigl|I_{t_1}+ G_{21}^{\hermitian}G_{21}K_{1|2}(I_{t_1}+ G_{21}^{\hermitian}G_{21}K_{1|2})^{-1}\bigr|\notag\\
	&=\log\frac{|I_{t_1}+ 2G_{21}^{\hermitian}G_{21}K_{1|2}|}
			  {|I_{t_1}+  G_{21}^{\hermitian}G_{21}K_{1|2}|}\notag\\
	&\le \min(t_1,r_2).
\end{align*}

\subsection{Noncoherent Partial Decode--Forward (Proof of the First Statement 
of Proposition~\ref{prop:gap-npdf})}\label{sec:npdf}
We use the following fact.

\begin{lemma}\label{lemm:covariance}
Let $K\succeq 0$ be of the form in \eqref{eq:covariance}.
Then, for every  $G_{31}$ and $G_{32}$,
we have 
	\begin{equation} \label{eq:cross-covariance}
	G_{ 31}K_1G_{ 31}^{\hermitian} +  G_{ 32}K_2G_{ 32}^{\hermitian} 
	\succeq  G_{ 32}K_{12}^{\hermitian}G_{ 31}^{\hermitian} + G_{ 31}K_{12}G_{ 32}^{\hermitian} .
	\end{equation}
\end{lemma}

\begin{IEEEproof}[Proof of Lemma~\ref{lemm:covariance}]
Consider
\[
\begin{bmatrix}
G_{ 31} &G_{ 32}
\end{bmatrix}
\begin{bmatrix} 
K_{1} &-K_{12} \\
-K_{12}^{\hermitian} & K_{2}
\end{bmatrix}
\begin{bmatrix}
G_{ 31} &G_{ 32}
\end{bmatrix}^{\hermitian}\succeq 0.
\]
\end{IEEEproof}

To prove Proposition~\ref{prop:gap-npdf}, let 
$K \succeq 0$ be of the form in~\eqref{eq:covariance}.
Let $\Xv_1\sim\mathrm{CN}(0,K_1)$ and $\Xv_2\sim\mathrm{CN}(0,K_2)$ be independent of each other,
and define $\Uv$ as in~\eqref{eq:uv-choice}.
Then, by Lemmas~\ref{lemm:det-inequality} and~\ref{lemm:covariance},
the first term of the minimum in the noncoherent partial decode--forward lower bound in~\eqref{eq:npdf-1} is
\begin{align}
&I(\Xv_1, \Xv_2; \Yv_3) \notag\\
&= \log | I_{r_3}+G_{31}K_1 G_{31}^{\hermitian} +  G_{32}K_2 G_{32}^{\hermitian} | \label{eq:npdf-bd} \\
	&\ge \log\Bigl | I_{r_3}   +  \half( G_{31}K_1 G_{31}^{\hermitian}  +  G_{32}K_2 G_{32}^{\hermitian} \notag\\[-0.5em]
	&\hphantom{\ge \log\Bigl | I_{r_3}   +  \half( G}
	+ G_{32}K_{12}^{\hermitian}G_{31}^{\hermitian}  + G_{31}K_{12}G_{32}^{\hermitian} )\Bigr | \label{eq:npdf-bd-0} \\
&\ge  \log | I_{r_3}  + G_{3\mystar} K G_{3\mystar}^{\hermitian} | 
- \min(t_1  +  t_2,\, r_3). \label{eq:npdf-bd-1}
\end{align}
Following steps similar to the coherent case in Section \ref{sec:co-pdf}, 
we have 
\begin{align*}
\Cov(\Xv_1 | \Uv, \Xv_2) 
&= \Cov(\Xv_1 | \Uv)  \\
&= K_1 (I_{t_1} + G_{21}^{\hermitian}G_{21} K_1)^{-1}
\end{align*}
and
\begin{align}
&I(\Uv;\Yv_2|\Xv_2)+I(\Xv_1;\Yv_3|\Xv_2,\Uv) \notag \\
&=\log\frac{|I_{r_3}+G_{ 31}\Cov(\Xv_1 | \Uv) G_{ 31}^{\hermitian}|}
	{|I_{r_2}+G_{ 21}\Cov(\Xv_1 | \Uv) G_{ 21}^{\hermitian}|} 	  \notag \\
&\qquad+\log{ |I_{r_2}+G_{ 21}K_{1}G_{ 21}^{\hermitian}|} \notag\\
&=\log|I_{t_1}+(G_{21}^{\hermitian}G_{21}+G_{31}^{\hermitian}G_{31})K_{1 }|  \notag \\
&\qquad+\log  \frac{ |I_{r_2}   +   G_{ 21}K_{1 }G_{ 21}^{\hermitian}|}
	         {|I_{t_1} + 2 G_{ 21}^{\hermitian}G_{ 21}K_{1 }|} \notag \\
&\ge \log|I_{t_1}+(G_{21}^{\hermitian}G_{21}+G_{31}^{\hermitian}G_{31})K_{1 }|
 - \min(t_1,r_2).
			 \label{eq:npdf-bd-2}
\end{align}
Comparing \eqref{eq:npdf-bd-1} and~\eqref{eq:npdf-bd-2} with the cutset bound in~\eqref{eq:cs-3}
completes the proof.

\subsection{Multiplicative Gap (Proofs of Theorem \ref{thm:multiplicative-gap-pdf}
and the Second Statement of Proposition~\ref{prop:gap-npdf}) }\label{sec:multiplicative-gap-pdf}

By setting $\Uv = \emptyset$ or $\Xv_1$ in \eqref{eq:npdf-1} 
and specializing~\eqref{eq:df-2} to independent $(X_1,X_2)$,
it can be readily checked that $\Rnpdf$ and $\max(\Rdf, \Rdt)$ are simultaneously lower bounded by
\begin{align}
	&\max\Bigl\{ {} 
	\max_{K_1,K_2} \min \bigl( 
	\log | I_{r_3}+G_{31}K_1 G_{31}^{\hermitian} +  G_{32}K_2 G_{32}^{\hermitian}|, \notag\\[-0.5em]
	&\hphantom{\max\Bigl\{ {} \max_{K_1,K_2} \min \bigl(}
	\log  | I_{r_2}  +  G_{21} K_{1} G_{21}^{\hermitian} |\bigr),\notag\\
	&\hphantom{\max\Bigl\{ {} }
	\;\, \max_{K_1} \log |I_{r_3}  +  G_{31} K_1 G_{31}^{\hermitian}  |\Bigr\}\notag\\
	&=\max_{K_1,K_2}\min \Bigl\{ {} 
	\log | I_{r_3}+G_{31}K_1 G_{31}^{\hermitian} +  G_{32}K_2 G_{32}^{\hermitian}|, \notag\\[-0.25em]
	&\hphantom{\;\mathrel{\ge}\max_{K_1,K_2}\min\Bigl\{ {} }
	\max\bigr(\log  | I_{r_2}  +  G_{21} K_{1} G_{21}^{\hermitian} |,\notag\\[-0.25em]
	&\hphantom{\;\mathrel{\ge}\max_{K_1,K_2}\min\Bigl\{ {} \max\bigr(}
	 \log |I_{r_3}  +  G_{31} K_1 G_{31}^{\hermitian}|\bigr)\Bigr\}.\label{eq:multiplicative-lb}
\end{align}
We further lower bound each term in~\eqref{eq:multiplicative-lb}. By
\eqref{eq:npdf-bd-0},
\begin{align}
&\log | I_{r_3}+G_{31}K_1 G_{31}^{\hermitian} +  G_{32}K_2 G_{32}^{\hermitian}|\notag\\
&\ge \log\Bigl | I_{r_3}   +  \half( G_{31}K_1 G_{31}^{\hermitian}  +  G_{32}K_2 G_{32}^{\hermitian} \notag\\[-0.5em]
	&\qquad\qquad\qquad
	+ G_{32}K_{12}^{\hermitian}G_{31}^{\hermitian}  + G_{31}K_{12}G_{32}^{\hermitian} )\Bigr | \notag\\
&\ge \half \log | I_{r_3}   +  ( G_{31}K_1 G_{31}^{\hermitian}  +  G_{32}K_2 G_{32}^{\hermitian} \notag\\[-0.25em]
&\qquad\qquad\qquad
	+ G_{32}K_{12}^{\hermitian}G_{31}^{\hermitian}  + G_{31}K_{12}G_{32}^{\hermitian} ) |. \label{eq:multiplicative-lb-1}
\end{align}
Similarly,
\begin{align}
&\max \big\{\log|I_{t_1}+ G_{21}^{\hermitian}G_{21}K_{1}|,
	\log|I_{t_1}+ \!G_{31}^{\hermitian}G_{31}K_{1}|\big\}
\notag\\
&\ge \half\big(\log|I_{t_1}+ \!G_{21}^{\hermitian}G_{21}K_{1}|
					+\log|I_{t_1}+ \!G_{31}^{\hermitian}G_{31}K_{1}|\big)
\notag\\
&\ge \half\log\big|I_{t_1}+ (G_{21}^{\hermitian}G_{21}+G_{31}^{\hermitian}G_{31})K_{1}\big|. \label{eq:multiplicative-lb-2}
\end{align}
Comparing~\eqref{eq:multiplicative-lb-1} and \eqref{eq:multiplicative-lb-2} with the cutset bound in \eqref{eq:cs-3} establishes that
\begin{align}
\Rpdf&\ge \max(\Rnpdf,\Rdf,\Rdt) \notag\\
&\ge \min\{\Rnpdf, \max(\Rdf,\Rdt)\} \ge \half\Rcs. \label{eq:pdf-lb}
\end{align}

\subsection{Decode--Forward and Direct Transmission (Proof of Proposition~\ref{prop:gap-df})}\label{Proof_prop-gap-df}
Consider the MIMO relay channel with 
$G_{31}=\text{diag}(g,1)$, $G_{21}=\text{diag}(1,g)$, $G_{32}=\text{diag}(g,g)$, $g> 1$, which 
is equivalent to a product of two mismatched single-antenna relay channels, one with
the direct channel stronger than the sender-to-relay channel and the other in the opposite direction.
Set $K_{1|2} = K_1 = K_2 = (P/2) I_2$ in~\eqref{eq:pdf-bd-1} and \eqref{eq:pdf-bd-3}, we have 
\begin{align}\label{eq:pdf-parallel}
	\Rpdf
	&\ge\min \bigg\{ \log \big(1 + g^2P \big)\Big(1 + (1+g^2)\frac{P}{2} \Big), \notag \\[-0.3em]
	&\hphantom{\,\,\ge\min \bigg\{ }
	\log \Big(1 + (1+g^2)\frac{P}{2} \Big)^2 -2\bigg\} \notag\\
	&=\log \Big(1 + (1+g^2)\frac{P}{2} \Big)^2 -2.
\end{align}
In comparison, 
\begin{align}
\Rdf = \Rdt
&= \max_{P_1+P_2\le P}\log (1 + P_1 )(1 + g^2P_2 ) \notag\\
&\le \log (1 + P )(1 + g^2P ).\label{eq:df-dt}
\end{align}
Therefore, we have
\begin{align*}
\Rpdf-\max(\Rdf,\Rdt)
	& \ge  \log \frac{\left(1 + (1+g^2)\frac{P}{2}\right)^2}{(1+P)(1+g^2 P)} - 2
\end{align*}
which tends to infinity as $g \to \infty$. 
Based on this example, 
more examples of larger dimensions
can be constructed.

\section{Compress--Forward}\label{sec:cf}

We prove Theorem~\ref{thm:gap-cf}.
Let $K \succeq 0$ be of the form in~\eqref{eq:covariance}.
Let $\Xv_1\sim\mathrm{CN}(0,K_1)$ and $\Xv_2\sim\mathrm{CN}(0,K_2)$ be independent of each other,
and 
\begin{equation} \label{eq:yv-choice}
\hat\Yv_2 =\Yv_2 + \hat\Zv_2 
\end{equation}
where $\hat\Zv_2\sim\mathrm{CN}(0,\sigma^2 I_{r_2})$ is
independent of $\Xv_1,$ $\Xv_2,$ $\Zv_2,$ and $\Zv_3$. 
Then, 
\begin{align}
&I(\Yv_2;\hat\Yv_2|\Xv_1,\Xv_2,\Yv_3) \notag\\
&=h(\hat\Yv_2|\Xv_1,\Xv_2,\Yv_3)
-h(\hat\Yv_2|\Xv_1,\Xv_2,\Yv_2,\Yv_3) \notag\\
&= r_2\log (1 + 1/\sigma^2) \label{eq:cf-bd2}
\end{align}
and
\begin{align}
&I(\Xv_1;\hat\Yv_2,\Yv_3|\Xv_2) \notag\\
&=  \log \frac{\Biggl | 	
	\begin{bmatrix}
	(1 + \sigma^2) I_{ r_2}     &   0\\
	0				  &   I_{ r_3}
	\end{bmatrix} +
	G_{\mystar 1} K_1 G_{\mystar 1}^{\hermitian} \Biggr|}
	{ \Biggl | 	\begin{bmatrix}
	(1 + \sigma^2) I_{ r_2}     &   0\\
	0				  &   I_{ r_3}
	\end{bmatrix} \Biggr| } 	 \notag \\
&=\log\Bigl | I_{t_1}+\Bigl (\frac{1}{1+\sigma^2}G_{21}^{\hermitian}G_{21}+G_{31}^{\hermitian} G_{31}\Bigr )K_1 \Bigr |   \label{eq:cf-bd3-1}\\
&\ge\log|I_{t_1}+ (G_{21}^{\hermitian}G_{21}+G_{31}^{\hermitian} G_{31})K_{1}  |\notag\\
&\qquad-\min(t_1,r_2+r_3)\log(1+\sigma^2). \label{eq:cf-bd3-2}
\end{align}
The first statement of Theorem~\ref{thm:gap-cf} is now established by
substituting~\eqref{eq:npdf-bd-1}, \eqref{eq:cf-bd2}, and \eqref{eq:cf-bd3-2} in the compress--forward lower bound in \eqref{eq:cf-2} 
and comparing it with the cutset bound in~\eqref{eq:cs-2}.
Setting $\sigma^2 = 1$ in \eqref{eq:cf-gap1} yields the second statement in \eqref{eq:cf-gap2}.

\section{Computation of the Capacity Bounds}
\label{sec:computation-capacity}

 \subsection{Formulations of Optimization Problems}
\subsubsection{Cutset Bound}\label{sec:compu-cutset}
Computing the cutset upper bound in \eqref{eq:cs-2} can be formulated as the following convex optimization 
problem \cite{Ng--Foschini2011}:
\begin{equation}
\begin{split}\label{eq:cvx}
	\text{maximize~~~}    &\Rcs\\
	\text{over~~~}	        &\Rcs\ge 0, K\succeq 0, K_{1|2}\succeq 0\\
	\text{subject to~~~}   &\Rcs\le   \log\bigr |I_{r_3}  +  G_{3\mystar} K G_{3\mystar}^{\hermitian} \bigr | \\
						 & \Rcs\le  \log \bigl | I_{r_2+r_3}  +  G_{\mystar1} K_{1|2} G_{\mystar1}^{\hermitian} \bigr|	\\
						 &\tr(A_1^{\hermitian} K A_1)\le P, \tr(A_2^{\hermitian} K A_2)\le P \\
						 & K- A_1 K_{1|2} A_1^{\hermitian} \succeq 0 
\end{split}
\end{equation}
where 
\[
A_1=\begin{bmatrix}
	I_{t_1}\\
	0_{t_2\times t_1}
	\end{bmatrix} \text{~~and~~}
	A_2=\begin{bmatrix}
	0_{t_1\times t_2}\\
	I_{t_2}
	\end{bmatrix}.
\]
The optimization problem in \eqref{eq:cvx} can be solved by standard convex optimization techniques or
packages, e.g., \cite{Grant--Boyd2013}.

\subsubsection{Partial Decode--Forward Lower Bound}

Since direct computation of~\eqref{eq:pdf-1} or~\eqref{eq:pdf-2} is intractable,
we instead consider three lower bounds on 
$\Rpdf$, namely, $\Rdf, \Rdt$, and the special case of $\Rpdf$ evaluated by~\eqref{eq:uv-choice},
and take the maximum of the three.
Note that all three lower bounds can be viewed as
the partial decode--forward lower bound evaluated by~\eqref{eq:uv-choice} with
a more general choice of $\Zv_2' \sim \mathrm{CN}(0, \sigma^2 I_{r_2})$,
where $\sigma^2 = \infty, 0, 1$, respectively. Considering more values of $\sigma^2$
can further improve the bound at the cost of complexity.

As for the cutset bound, 
both $\Rdf$ and $\Rdt$ can be computed efficiently as a convex optimization problem.
The third bound, characterized by~\eqref{eq:pdf-bd-1} and~\eqref{eq:pdf-bd-2},
is nonconvex. Thus, we evaluate the bound with the optimal solution to
the convex optimization problem defined by~\eqref{eq:pdf-bd-1} and~\eqref{eq:pdf-bd-3}.
A similar approach can be taken for computation of $\Rnpdf$.

\subsubsection{Compress--Forward Lower Bound}
We consider two convex lower bounds on $\Rcf$,
namely, the special case of $\Rcf$ evaluated by~\eqref{eq:yv-choice}
with $\sigma^2 = 1$, namely,
\begin{align*}
&{\max_{K_1,K_2}\min\Big\{ {} }
		\log | I_{r_3}+G_{31}K_1 G_{31}^{\hermitian} +  G_{32}K_2 G_{32}^{\hermitian} |
		-r_2,\notag\\[-0.5em]
&\hphantom{\max_{K_1,K_2}\min\Big\{ {} }
		\log\Bigl | I_{t_1}+\Bigl (\frac{1}{2}G_{21}^{\hermitian}G_{21}+G_{31}^{\hermitian} G_{31}\Bigr )K_1 \Bigr |			
		\Big\}
\end{align*}
and $\Rdt$ (which corresponds to $\sigma^2 = \infty$).
As in the case of partial decode--forward, considering more values of $\sigma^2$
can further improve the bound at the cost of complexity.

\subsection{Numerical Results}

\begin{figure}[h] \center
    \psfig{figure=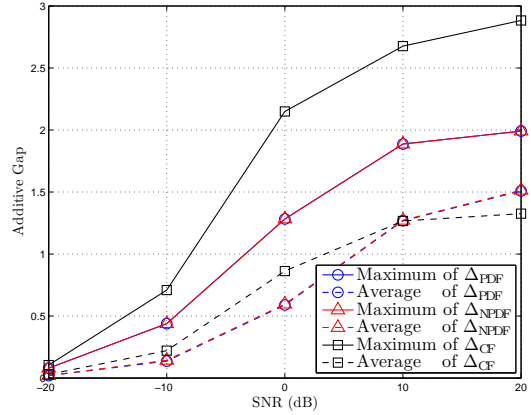, width=7cm}
    \caption{The additive gaps between the cutset bound and the  partial decode--forward and compress--forward lower bounds for randomly generated
    $2\times 2$ MIMO relay channels.} \label{CS_22MIMO_complex_gamma_additive}
\end{figure}

\begin{figure}[h] \center
    \psfig{figure=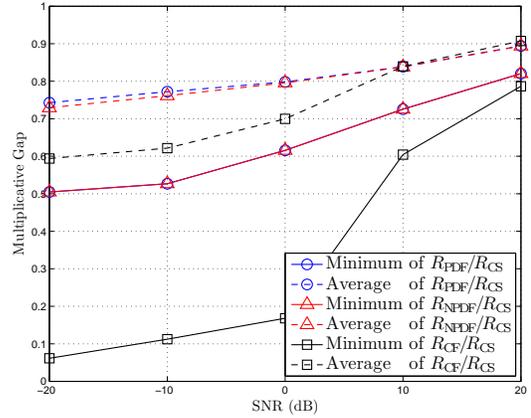, width=7cm}
    \caption{The multiplicative gaps between the cutset bound and the  partial decode--forward and compress--forward lower bounds for randomly generated
    $2\times 2$ MIMO relay channels.} \label{CS_22MIMO_complex_gamma_multiplicative}
\end{figure}

We consider the additive and multiplicative gaps on 2000 
$2\times 2$ MIMO relay channels with
random channel gains independently distributed according to $\mathrm{CN}(0,1)$.
The gaps are evaluated by relaxed bounds discussed in the previous subsection. 
The maximum and average of the additive gaps are shown in  
Fig.~\ref{CS_22MIMO_complex_gamma_additive} and similar multiplicative gaps 
are shown in Fig.~\ref{CS_22MIMO_complex_gamma_multiplicative}. 
The simulation results are consistent with the theoretical predictions in 
Theorems~\ref{thm:gap-pdf},~\ref{thm:multiplicative-gap-pdf}, and~\ref{thm:gap-cf}, and 
Proposition~\ref{prop:gap-npdf}.


\section{Half-Duplex MIMO Relay Channels}\label{sec:half-duplex}

\begin{figure}[h]
\begin{center}
\subfigure[Sender frequency division.]{
\smallskip
 \centering
 \small
 \psfrag{xr}[r]{$\Xv_1''$}
 \psfrag{xd}[r]{$\Xv_1'$}
 \psfrag{a}[b]{$G_{21}$ }
 \psfrag{z1}[b]{$\Zv_2$}
 \psfrag{y1}[l]{$\Yv_{\!2}$}
 \psfrag{x1}[r]{$\Xv_2$}
 \psfrag{x1}[r]{$:\hspace{-2pt}\Xv_2$\!}
 \psfrag{b}[b]{$G_{32}$}
 \psfrag{z}[b]{$\Zv_3$}
 \psfrag{y}[l]{$\Yv_3$}
 \psfrag{1}[t]{$G_{31}$ }
 \includegraphics[scale=0.45]{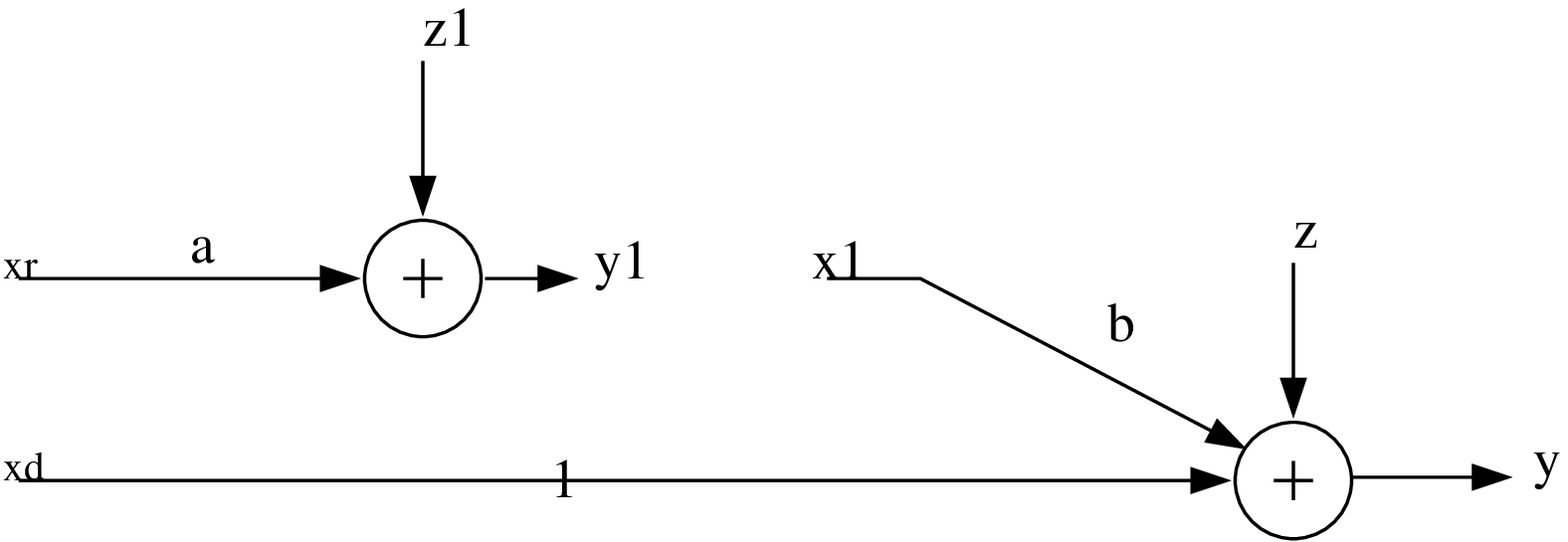}
 \label{fig:relay-sfd}}

\medskip  
\subfigure[Receiver frequency division.]{
 \centering
 \small
 \psfrag{x}[r]{$\Xv_1$}
 \psfrag{a}[b]{$G_{21}$}
 \psfrag{z1}[b]{$\Zv_2$}
 \psfrag{y1}[l]{$\Yv_{\!2}$}
 \psfrag{x1}[r]{$\Xv_2$}
 \psfrag{x1}[r]{$:\hspace{-2pt}\Xv_2$\!}
 \psfrag{b}[b]{$G_{32}$}
 \psfrag{zs}[t]{$\Zv_3'$}
 \psfrag{z'}[b]{$\Zv_3''$}
 \psfrag{ys}[l]{$\Yv_3'$}
 \psfrag{y'}[l]{$\Yv_3''$}
 \psfrag{1}[t]{$G_{31}$}
 \includegraphics[scale=0.42]{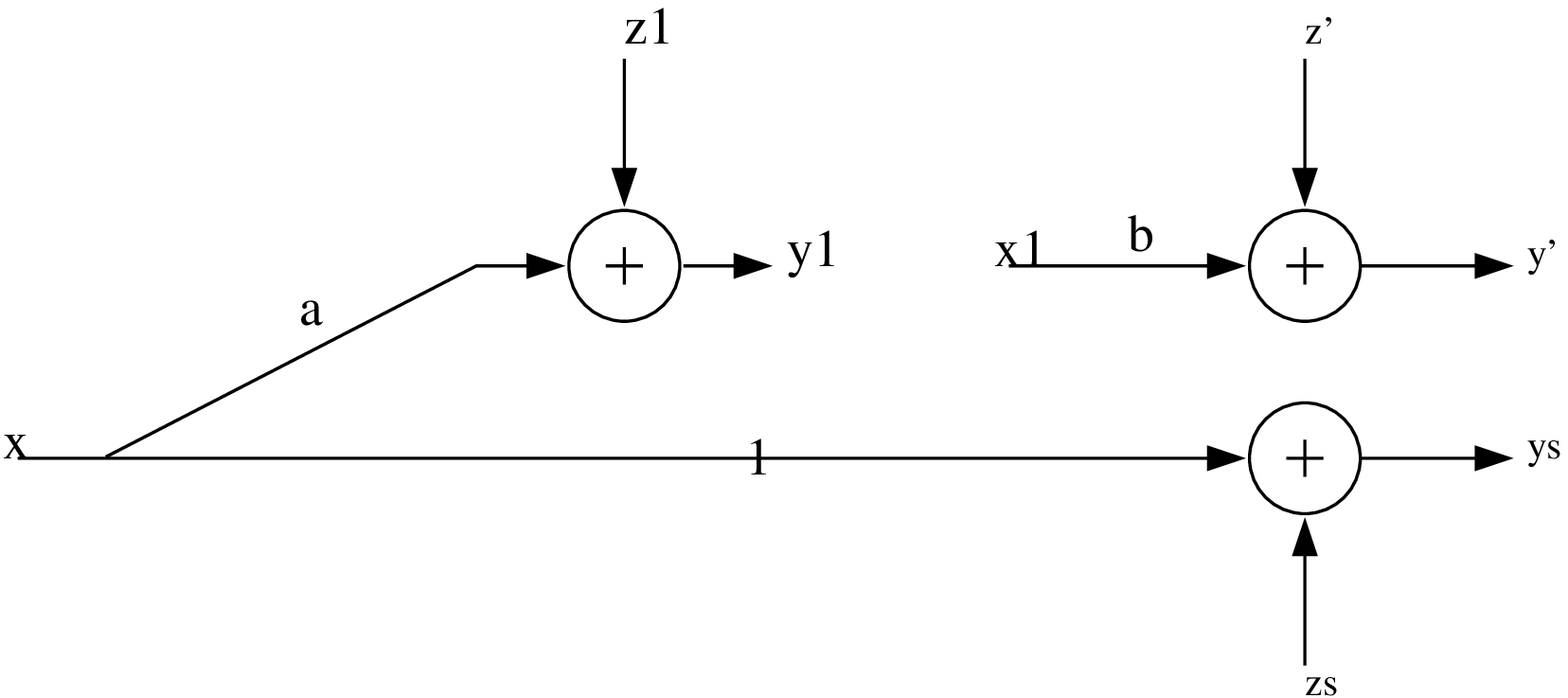}%
 \medskip
  \label{fig:relay-rfd}}
 \caption{Half-duplex MIMO relay channels.}
\end{center}
 \end{figure}

Half-duplex relay channel models are 
often investigated to study wireless communication systems in which relays cannot send and receive in the same time slot or frequency band.
There are two different types of half-duplex models. One is the sender frequency-division (SFD) MIMO relay channel (Fig.~\ref{fig:relay-sfd}), in which the channel from the sender to the relay, $\Xv_1'' \to  \Yv_2$, is orthogonal to the multiple access channel from the sender and the relay to the receiver, $(\Xv_1', \Xv_2) \to  \Yv_3$.
The other is the receiver frequency-division (RFD) MIMO relay channel (Fig.~\ref{fig:relay-rfd}), in which the channel $\Xv_2 \to  \Yv_3''$ is orthogonal to the broadcast channel $\Xv_1 \to  (\Yv_2, \Yv_3')$.
Both can be viewed as special cases of the general (full-duplex) MIMO relay channel model.
For example, the SFD model follows  
by setting 
$\begin{bmatrix} G_{31} & 0 \end{bmatrix}\in \mathbb{C}^{r_3\times (t_1'+t_1'')}$ and $\begin{bmatrix} 0 &G_{21} \end{bmatrix}\in \mathbb{C}^{r_3\times (t_1'+t_1'')}$
in \eqref{eq:channel-model}.
Consequently, our main results in Section~II continue to hold with $t_1 = t_1' + t_1''$ and $r_3 = r_3' + r_3''$
for the SFD and RFD cases, respectively.

In the following, we present tighter results that exploit the half-duplex channel structure.
The proofs are similar to the full-duplex case in basic analysis techniques and relegated to the Appendix.

\subsection{Sender Frequency-Division MIMO Relay Channels}
It has been shown by El~Gamal
and Zahedi \cite{El-Gamal--Zahedi2005}
that
the relay channel capacity is achieved by partial decode--forward
when the sender has orthogonal components.
We specialize this result to the multiple-antenna case.
 
\begin{proposition} \label{Prop:C-sfd}
The capacity of the SFD MIMO relay channel is 
\begin{align} \label{eq:C-sfd}
	C &= \Rcs 
	= \Rpdf \notag\\
	&=\sup_{F(\xv_1',\xv_2)F(\xv''_1)}\min\{ {} I(\Xv'_1,\Xv_2;\Yv_3),\notag\\[-0.5em]
	&\hphantom{\;\mathrel{=}\sup_{F(\xv_1',\xv_2)F(\xv''_1)}\min\{ {} }
	I(\Xv''_1;\Yv_2)+I(\Xv'_1;\Yv_3|\Xv_2)\}\notag\\
     &=\max_{K}\min\bigg\{ {}
     \log\bigg|I_{ r_3}+ G_{3\mystar}
			\begin{bmatrix}
  			 K'_{1} &  K'_{12}\\
  			 (K'_{12})^{\hermitian} & K_{2}
 			\end{bmatrix}G_{3\mystar}^{\hermitian}\bigg|,\notag\\
	&\hphantom{\;\mathrel{=}\max_{K}\min\bigg\{ {} }
	  \log|I_{ r_2}+ G_{21}K_{1}''G_{21}^{\hermitian}| \notag\\[-0.3em]
	&\hphantom{\quad=\max_{K}\min\bigg\{}
	  +\log|I_{ r_3}+ G_{31}K'_{1|2}G_{31}^{\hermitian}|\bigg\} 
\end{align}
where the supremum is over all $F(\xv_1', \xv_2)F(\xv''_1)$ such that $\E((\Xv'_1)^{\hermitian}\Xv'_1)+\E((\Xv''_1)^{\hermitian}\Xv''_1) \le P$ and $\E(\Xv_2^{\hermitian}\Xv_2)\le P$,
the maximum is over all $(t_1'+t_1''+t_2) \times (t_1'+t_1''+t_2) $ matrices
\begin{equation} \label{eq:covariance2}
K = 
\begin{bmatrix}
K'_1		&0    & K'_{12} \\
0   		&K''_1& 0 \\
(K'_{12})^{\hermitian}& 0 & K_2
\end{bmatrix} \succeq 0
\end{equation}
such that
$\tr(K'_1+K''_1) \le P$, $\tr(K_2) \le P$, 
and $K'_{1|2} = K'_1 - K'_{12} K_2^{-1} (K'_{12})^{\hermitian}$.
\end{proposition}

We can establish the following (in addition to obvious corollaries from the full-duplex results).

\begin{proposition} \label{Prop:gap-pdf-sfd}
For every $G_{21}$, $G_{31}$, $G_{32}$, and $P$,
\begin{align}
	\Dnpdf =C-\Rnpdf&\le \min(t'_1+t_2,\, r_3)\label{eq:gap-pdf-sfd}.
	\end{align}
\end{proposition}

\medskip

\begin{proposition} \label{Prop:gap-cf-sfd}
For every $G_{21}$, $G_{31}$, $G_{32}$, and $P$,
\begin{align}
	\Dcf=C-\Rcf&\le \min(t'_1+t_2,\, r_3)+r_2.\label{eq:gap-cf-sfd}
\end{align}

\end{proposition}
\medskip

\subsection{Receiver Frequency-Division MIMO Relay Channels}

The capacity in this case is not known in general.
 
\begin{proposition}
The capacity $C$ of the RFD MIMO relay channel is upper bounded by
\begin{align}
\Rcs   &=\sup_{F(\xv_1)F(\xv_2)}\min\{ {}
		I(\Xv_1;\Yv'_3)+I(\Xv_2;\Yv''_3),\notag\\[-0.5em]
	  &\hphantom{\;\mathrel{=}\sup_{F(\xv_1)F(\xv_2)}\min\{ {} } 
	  I(\Xv_1;\Yv_2,\Yv'_3)\}
\notag\\
	  &=\max_{K_1} \min\Bigl\{{}
	  \log|I_{ r_3}+G_{31}K_{1}G_{31}^{\hermitian}|\notag\\[-0.5em]
	  &\hphantom{\;\mathrel{=}\max_{K_1} \min\Bigl\{{}}
	  \qquad +\max_{K_2}\log|I_{ r_3}+G_{32}K_{2}G_{32}^{\hermitian}|,\notag\\[-0.3em]
	  &\hphantom{\;\mathrel{=}\max_{K_1} \min\Bigl\{{}}
	  \log|I_{ t_1}+ (G_{21}^{\hermitian}G_{21}+G_{31}^{\hermitian}G_{31})K_{1}|\Bigr\} \label{eq:cs-rfd}
\end{align}
where the supremum is over all $F(\xv_1)F(\xv_2)$ such that $\E(\Xv_j^{\hermitian}\Xv_j) \le P$, $j = 1,2$, and
the maxima are over all $K_1, K_2 \succeq 0$ such that $\tr(K_j)\le P$, $j=1,2$.
\end{proposition}

As in the cutset bound, coherent transmission is irrelevant.
\begin{proposition}
For every $G_{21}, G_{31}, G_{32}$, and $P$, 
\[
\Rpdf = \Rnpdf
\]
and consequently
\[
\Dpdf = \Dnpdf \le\min(t_1, r_2).
\]
\end{proposition}

We can further establish the following.

\begin{proposition} \label{Prop:gap-cf-rfd}
For every $G_{21}, G_{31}, G_{32}$, and $P$, 
\begin{align}
	 &\Dcf \le\max [ \min(t_1,\, r_2 + r'_3) , r_2 ]. \label{eq:gap-cf-rfd}
\end{align}
\end{proposition}

\appendix

\begin{IEEEproof}[Proof of Proposition \ref{Prop:gap-pdf-sfd}]
Set $K'_{12}=0$ and $K_{1|2}' = K_1'$ in \eqref{eq:C-sfd}. The result follows by similar arguments to~\eqref{eq:npdf-bd-1}. 
\end{IEEEproof}

\begin{IEEEproof}[Proof of Proposition \ref{Prop:gap-cf-sfd}]
Let $K$ be the form of \eqref{eq:covariance2}. Let $\Xv'_1\sim \mathrm{CN}(0,K'_1)$, $\Xv''_1\sim \mathrm{CN}(0,K''_1)$, and $\Xv_2\sim \mathrm{CN}(0,K_2)$ be independent and $\hat\Yv_2 =\Yv_2 + \hat\Zv_2$, where $\hat\Zv_2\sim\mathrm{CN}(0, I_{r_2})$ is
independent of $(\Xv'_1, \Xv''_1, \Xv_2, \Zv_2, \Zv_3)$. 
Then, 
\begin{align*}
	&I(\Xv_1 , \Xv_2; \Yv_3) - I(\Yv_2;\hat{\Yv}_2|\Xv_1 ,\Xv_2,\Yv_3)\notag\\
	&\qquad=\log|I_{ r_3}+G_{31}K'_1G_{31}^{\hermitian}+G_{32}K_2G_{32}^{\hermitian}|
	-r_2
\end{align*}
and
\begin{align*}
&I(\Xv_1;\hat{\Yv}_2 ,\Yv_3|\Xv_2)\notag\\
&=\log|I_{ r_3}+G_{31}K'_1G_{31}^{\hermitian}|
   +\log|I_{ r_2}+(1/2)G_{21}K''_1G_{21}^{\hermitian}| \notag\\
&\ge \log|I_{ r_3}+G_{31}K'_1G_{31}^{\hermitian}|
   +\log|I_{ r_2}+G_{21}K''_1G_{21}^{\hermitian}| \notag\\
&\qquad   - \min(t_1'',r_2).
\end{align*}
The gap due to the first term, $\min(t_1'+t_2,r_3) + r_2$, 
follows by similar arguments to the proof of Proposition~\ref{Prop:gap-pdf-sfd},
which dominates the gap due to the second term. 
\end{IEEEproof}

\begin{IEEEproof}[Proof of Proposition \ref{Prop:gap-cf-rfd}]
The compress--forward lower bound in \eqref{eq:cf-2} simplifies to 
\begin{align}
\Rcf
&=\sup\min \bigl\{ {}
	I(\Xv_1; \Yv'_3) +I(\Xv_2;\Yv''_3)\!-\! I(\Yv_2;\hat{\Yv}_2|\Xv_1),\notag\\[-0.2em]
&\hphantom{\;\mathrel{=}\sup\min \bigl\{ {} }
	I(\Xv_1;\hat{\Yv}_2 ,\Yv'_3)\bigr \} \label{eq:cf-rfd}
\end{align}
where the supremum is over all conditional distributions $F(\xv_1)F(\xv_2)  F(\hat{\yv}_2 |\yv_2 )$
such that $\E(\Xv_j^{\hermitian}\Xv_j) \le P$, $j = 1,2$.
Let $\Xv_1\sim\mathrm{CN}(0,K_1)$ and $\Xv_2\sim\mathrm{CN}(0,K_2)$ be independent of each other. Let $\hat\Yv_2 =\Yv_2 + \hat\Zv_2$, where $\hat\Zv_2\sim\mathrm{CN}(0, I_{r_2})$ is
independent of $(\Xv_1, \Xv_2, \Zv_2, \Zv'_3, \Zv''_3)$. Then,
\begin{align*}
	&I(\Xv_1; \Yv'_3) +I(\Xv_2;\Yv''_3)- I(\Yv_2;\hat{\Yv}_2|\Xv_1)\notag\\
	&=\log|I_{r_3}+G_{31}K_1 G_{31}^{\hermitian}|
	+ \log|I_{r_3}+G_{32}K_2G_{32}^{\hermitian}| - r_2
\end{align*}
and
\begin{align*}
	&I(\Xv_1;\hat{\Yv}_2 ,\Yv'_3)
	=\log\Bigl|I_{t_1}+ \Bigl(\half G_{21}^{\hermitian}G_{21}+G_{31}^{\hermitian}G_{31}\Bigr)K_1\Bigr|.
\end{align*}
The rest of the proof follows similar steps to that of Proposition~\ref{Prop:gap-cf-sfd}.
\end{IEEEproof}


\newcommand{\noopsort}[1]{}

\end{document}